\documentclass[journal=acsanm,manuscript=article]{achemso}

\usepackage{caption}
\usepackage{xcolor}

\usepackage[labelfont=bf]{caption}
\usepackage[version=3]{mhchem} % Formula subscripts using \ce{}
\newcommand*{\citen}[1]{%
  \begingroup
    \romannumeral-`\x % remove space at the beginning of \setcitestyle
    \setcitestyle{numbers}%
    \cite{#1}%
  \endgroup   
}

\author{Gabriele Baglioni}
\affiliation[TU Delft]
{Kavli Institute of Nanoscience, Delft University of Technology, The Netherlands}
\email{G.Baglioni@tudelft.nl}
\author{Roberto Pezone}
\affiliation[TU Delft]
{Laboratory of Electronic Components, Technology and Materials, Delft University of Technology, The Netherlands}
\author{Sten Vollebregt}
\affiliation[TU Delft]
{Laboratory of Electronic Components, Technology and Materials, Delft University of Technology, The Netherlands}
\author{Katarina Cvetanović}
\affiliation[UBelgrade]{Center for Microelectronic Technologies, Institute of Chemistry, Technology and Metallurgy, University of Belgrade, Serbia}
\author{Marko Spasenović}
\affiliation[UBelgrade]{Center for Microelectronic Technologies, Institute of Chemistry, Technology and Metallurgy, University of Belgrade, Serbia}
\author{Dejan Todorović}
\affiliation{Dirigent Acoustics Ltd, Belgrade, Serbia}
\author{Hanqing Liu}
\affiliation[TU Delft]
{Department of Precision and Microsystems Engineering, Delft University of Technology, The Netherlands}
\author{Gerard J. Verbiest}
\affiliation[TU Delft]
{Department of Precision and Microsystems Engineering, Delft University of Technology, The Netherlands}
\author{Herre S.J. van der Zant}
\affiliation[TU Delft]
{Kavli Institute of Nanoscience, Delft University of Technology, The Netherlands}
\author{Peter G. Steeneken}
\affiliation[TU Delft]
{Kavli Institute of Nanoscience, Delft University of Technology, The Netherlands}
\alsoaffiliation[TU Delft]
{Department of Precision and Microsystems Engineering, Delft University of Technology, The Netherlands}
\email{P.G.Steeneken@tudelft.nl}

\title[Graphene microphones]
  {Ultra-sensitive graphene membranes for microphone applications}


\begin{document}

\begin{abstract}
Microphones exploit the motion of suspended membranes to detect sound waves. Since the microphone performance can be improved by reducing the thickness and mass of its sensing membrane, graphene-based microphones are expected to outperform state-of-the-art microelectromechanical (MEMS) microphones and allow further miniaturization of the device. Here, we present a laser vibrometry study of the acoustic response of suspended multilayer graphene membranes for microphone applications. We address performance parameters relevant for acoustic sensing, including mechanical sensitivity, limit of detection and nonlinear distortion, and discuss the trade-offs and limitations in the design of graphene microphones.  We demonstrate superior mechanical sensitivities of the graphene membranes, reaching more than 2 orders of magnitude higher compliances than commercial MEMS devices, and report a limit of detection as low as 15 dB$_\mathrm{SPL}$, which is $10-15$ dB lower than that featured by current MEMS microphones.
\end{abstract}


\section{Introduction}
MEMS microphone technology, based on Si manufacturing processes, has benefited from the proliferation of portable electronic devices, experiencing unprecedented market growth\cite{malcovati_evolution_2018} as well as continuous design and production improvements\cite{zawawi_review_2020,shah_design_2019}. One of the main device development targets is the optimization of the mechanical sensitivity, which determines the microphone's ability to pick up sound. The mechanical sensitivity, defined as the membrane's displacement amplitude per unit sound pressure, scales inversely with the thickness and stress of the sensing membrane\cite{zawawi_review_2020}. Complex fabrication techniques involving corrugated membranes\cite{corrugatedMEMS, corrugatedMEMS2} or not-fully supported membranes\cite{springMEMS,springMEMS2, springMEMS3} have been implemented to reduce residual fabrication stress  and boost sensitivity of thin MEMS membranes.

Being ultrathin and lightweight, suspended graphene membranes are excellent candidates for use in electrostatically actuated devices\cite{AbdelGhany_2016,Gswitch2014,cartamil-bueno_graphene_2018} and sensors\cite{peterR}, such as pressure sensors\cite{davP, berP, makaP}, gas sensors \cite{IrekGas} and accelerometers \cite{fan_suspended_2019} as well as microphones\cite{zhou_graphene_2015, wittmann_graphene_2019, todorovic_multilayer_2015,wood_design_2019,woo_realization_2017,xu_realization_2021,MohdMustapha2019CharacterizationOG, xu_realization_2020}. Thanks to their atomic thickness, graphene membranes could be made more than a factor 100-1000 times thinner than typical 0.1-1.0 $\mu$m thick MEMS membranes, resulting in a significant increase of the microphone mechanical sensitivity without requiring complex device structures. On top of that, graphene is an excellent conductor and thus requires no additional layer for electrical readout. %In addition, this advantage can lead to further miniaturization of the device and smaller production costs compared to Si-based technologies.
Previous studies have demonstrated the fabrication of microphones using graphene-based membranes either with multilayer graphene \cite{zhou_graphene_2015, wittmann_graphene_2019, todorovic_multilayer_2015} or with a composite structure made of bilayer or multilayer graphene and a thick ($>100$ nm) PMMA layer\cite{wood_design_2019,woo_realization_2017,xu_realization_2021,MohdMustapha2019CharacterizationOG, xu_realization_2020}. In general, these works focused on fabricating a condenser microphone structure, involving either wet or dry transfer \cite{todorovic_multilayer_2015,wood_design_2019,woo_realization_2017} of large graphene membranes (from 2 to 12 mm in diameter) over pre-patterned substrates or via dry etching of a sacrificial layer\cite{xu_realization_2021}. In these devices, the incoming sound is transduced to an electrical signal via the change in capacitance between a fixed backplate and the movable membrane. Although these works have demonstrated successful capacitive readout of audio signals with high output voltage per unit pressure, other important device performance parameters, like the mechanical sensitivity, the signal-to-noise ratio (SNR), total harmonic distortion (THD), bandwidth and dynamic range, have been less extensively characterized (see parameter definitions in Supplementary Information S1). 

In this work, we use a Laser Doppler Vibrometer (LDV) to carry out a detailed study of the response of multilayer graphene (MLG) membranes to acoustic actuation, and determine their most important performance parameters such that they can be compared to the state-of-the-art. The advantage of optical vibrometry is that it allows direct determination of the mechanical response of graphene membranes to sound, in contrast to electrical methods, where the output voltage depends on the specifics of the readout circuit. Moreover, by using it to characterize freestanding membranes, it allows measurement of the intrinsic membrane characteristics without including effects of a backplate that is used in capacitive condenser microphones. Thus, we gain deeper insights into graphene's acoustic properties, which is crucial for the design of future MEMS graphene microphones. 
\section{Experimental section}
\begin{figure}[H]
\includegraphics[width = \linewidth]{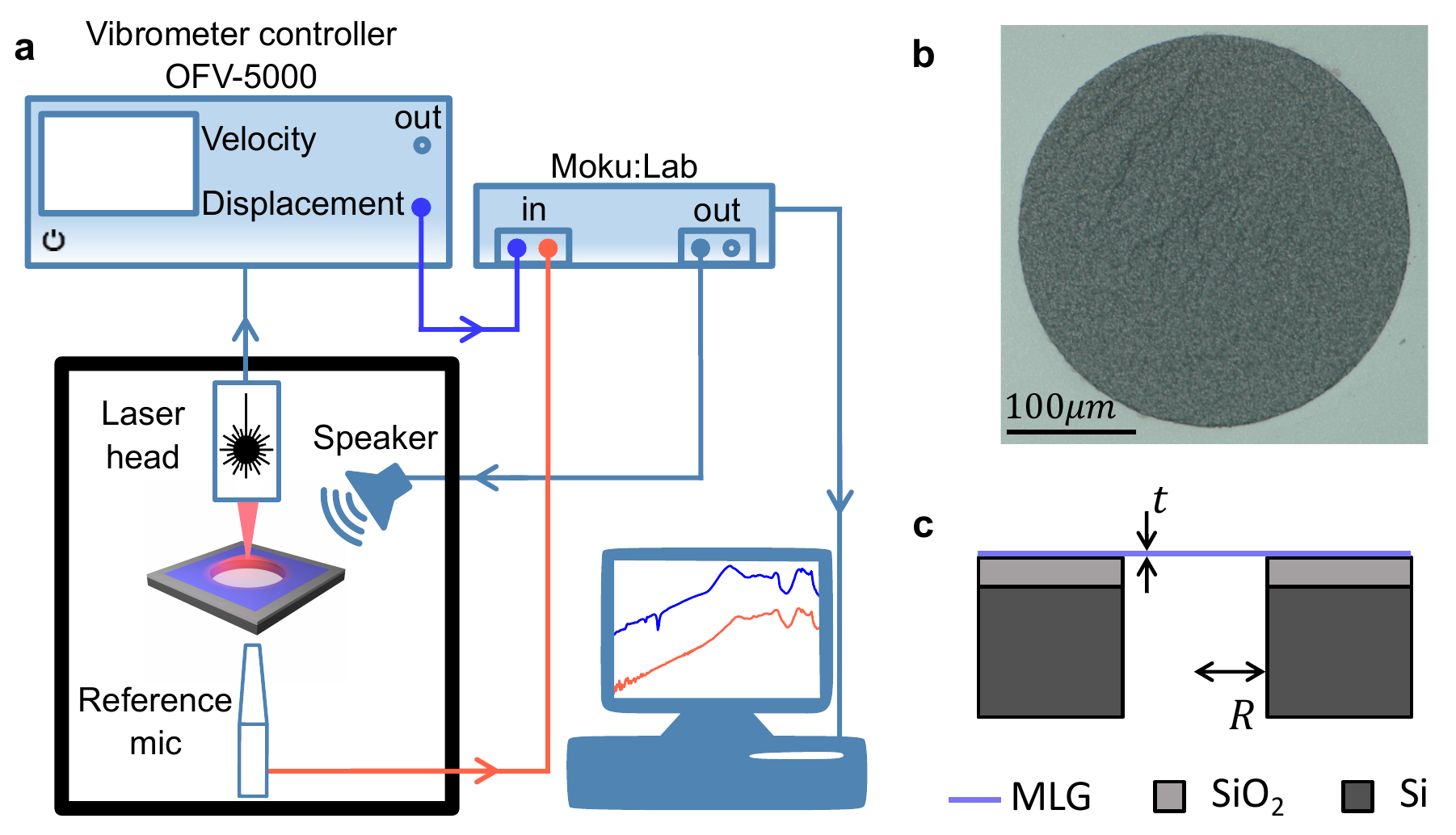}
\caption{\textbf{Experimental setup and samples.} \textbf{a} Schematics of the experimental setup. The vibrometer measures the dynamic motion of the graphene membrane as a result of the sound from a speaker at a distance of $\sim$ 2 cm, while a reference microphone, that is mounted within $\sim$ 5 mm below the chip, detects the sound level at the sample location. Measurements are controlled via the Moku:Lab using the spectrum analyzer and frequency response analyser functions. The setup is placed inside a sound proof box. \textbf{b-c} Optical image and schematic cross section of a multilayer graphene membrane (thickness $t \sim 8 $ nm) transferred over a through-hole with a diameter $d =2 R = 350 \, \mu$m in a Si/SiO$_2$ substrate.}
\label{fgr:Fig1}
\end{figure}

Figure \ref{fgr:Fig1}a shows the schematic of the setup used to characterize the multilayer graphene membranes. A single-point Laser Doppler Vibrometer LDV (OFV-5000 vibrometer controller and OFV-534 fiber-coupled vibrometer sensor head) is used to measure the displacement at the center of the membrane. A reference microphone (Sonarworks XREF20), placed below the sample under study, detects the input sound pressure level from a commercially available speaker used to acoustically actuate the graphene membrane. The displacement signal is reconstructed by the DD-900 decoder of the vibrometer controller with resposivity, $R_\text{LDV}$ set between 100 nm/V and 1 $\mu$m/V. The spectrum analyzer and frequency response analyser functions of Moku:Lab FPGA-based signal generator and analyzer are used to measure the displacement signal from the vibrometer and sound pressure signal from the microphone as well as controlling the driving signal to the speaker. A sound proof box encloses the setup to reduce influence of background noise. 

An optical picture of a typical graphene membrane and its schematic cross-section are shown in Fig. \ref{fgr:Fig1}b,c. The free-standing membranes are made of multilayer graphene with a thickness of $\sim 8$ nm grown on Si/SiO$_2$/Mo (50 nm) by Low-Pressure Chemical Vapor Deposition in an Aixtron Black Magic reactor at 1000$^{\circ}$C with H$_2$ - CH$_4$ as carbon precursor source. The Mo seed layer under the graphene is wet-etched with H$_2$O$_2$ and deionized water, after which the graphene remains on the Si/SiO$_2$ substrate \cite{Vollebregt2016}. The graphene is finally immersed in DI-water until it delaminates and it is carefully wet-transferred on a Si/SiO$_2$ substrate (thickness of $\sim 520 \,\mu$m) with pre-patterned holes. These holes, with a diameter of $\sim350-600 \,\mu$m were etched through the silicon chips by DRIE (Deep Reaction Ion Etching) and buffered oxide etch (BOE) to remove the SiO$_2$ hard mask. Finally, the chip with suspended graphene membranes is dried in atmospheric conditions for $>$ 10 hours. The crystallinity of the graphene as well as its thickness were investigated via Raman and atomic force microscopy (see Figure S1 and S2 in the Supplementary Information).

\begin{figure}[H]
\includegraphics[width = \linewidth]{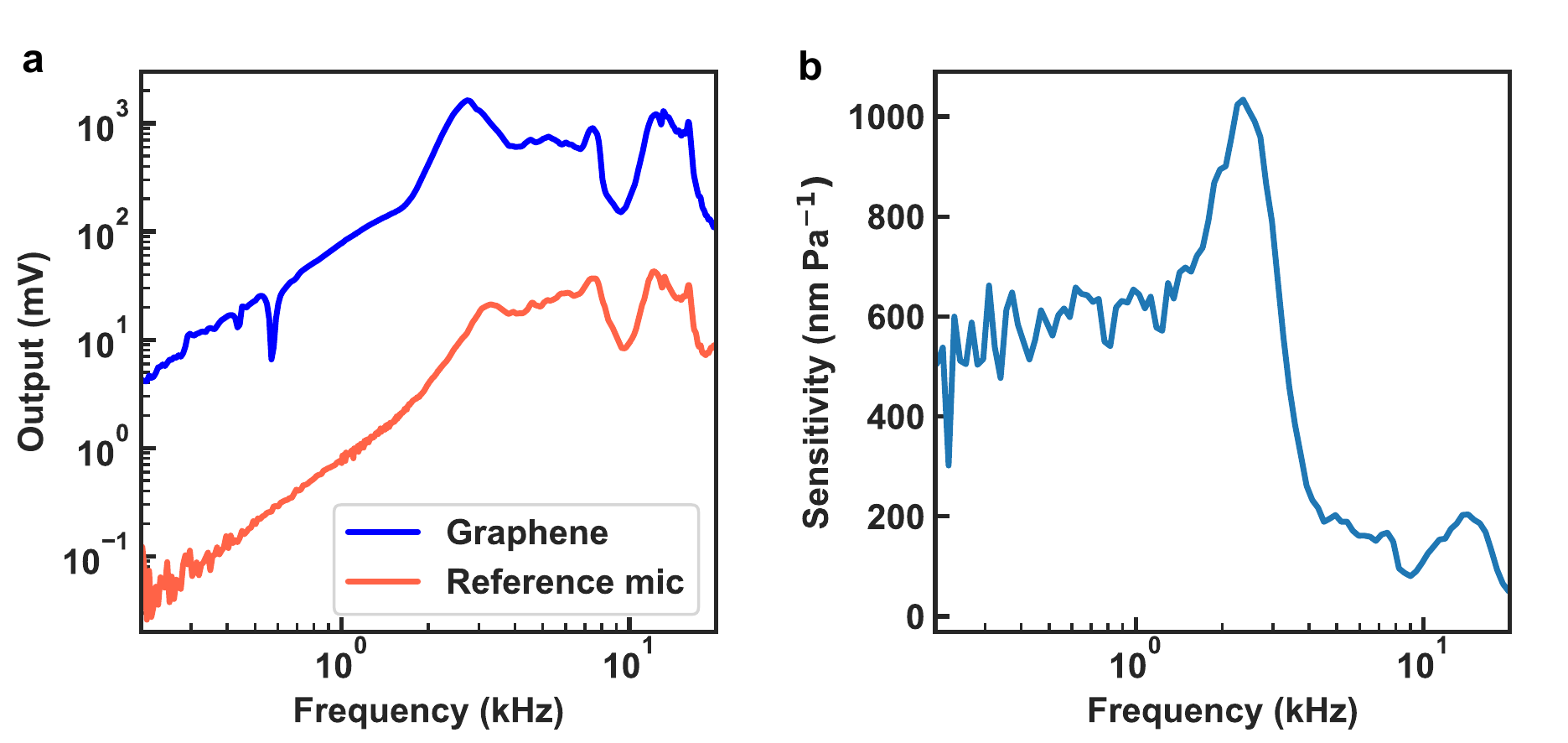}
\caption{\textbf{Frequency response measurements.} \textbf{a} LDV displacement data (blue line) of a graphene membrane (diameter $d = 350 \,\mu$m) and sound pressure data (light red line) recorded by the reference microphone between 200 Hz and 20 kHz. \textbf{b} Mechanical sensitivity  of the membrane, as determined from \textbf{a}.}
\label{fgr:Fig2}
\end{figure}

\section{Results}
We investigate the acoustic spectrum of the graphene membranes by measuring their center displacement in response to an acoustic chirp from 200 Hz to 20 kHz. Figure \ref{fgr:Fig2}a shows typical frequency responses  from the LDV (in blue) and from the reference microphone (in red). The vibrometer outputs a voltage signal $V_\text{LDV}$ proportional to the membranes displacement, $z$, like $V_\text{LDV} = zR_\text{LDV}$. Similarly, the output voltage of the reference microphone is proportional to the incoming sound pressure level by its calibrated sensitivity in the audible range. After correcting for the vibrometer's responsivity (in nm/V) and for the calibrated sensitivity of the microphone (in mV/Pa),  the ratio of the two voltage signals in Fig \ref{fgr:Fig2}a yields the mechanical sensitivity in nm/Pa of the graphene membrane, which is shown in Fig. \ref{fgr:Fig2}b.

\begin{figure}[H]
    \centering
    \includegraphics[width = 1\linewidth]{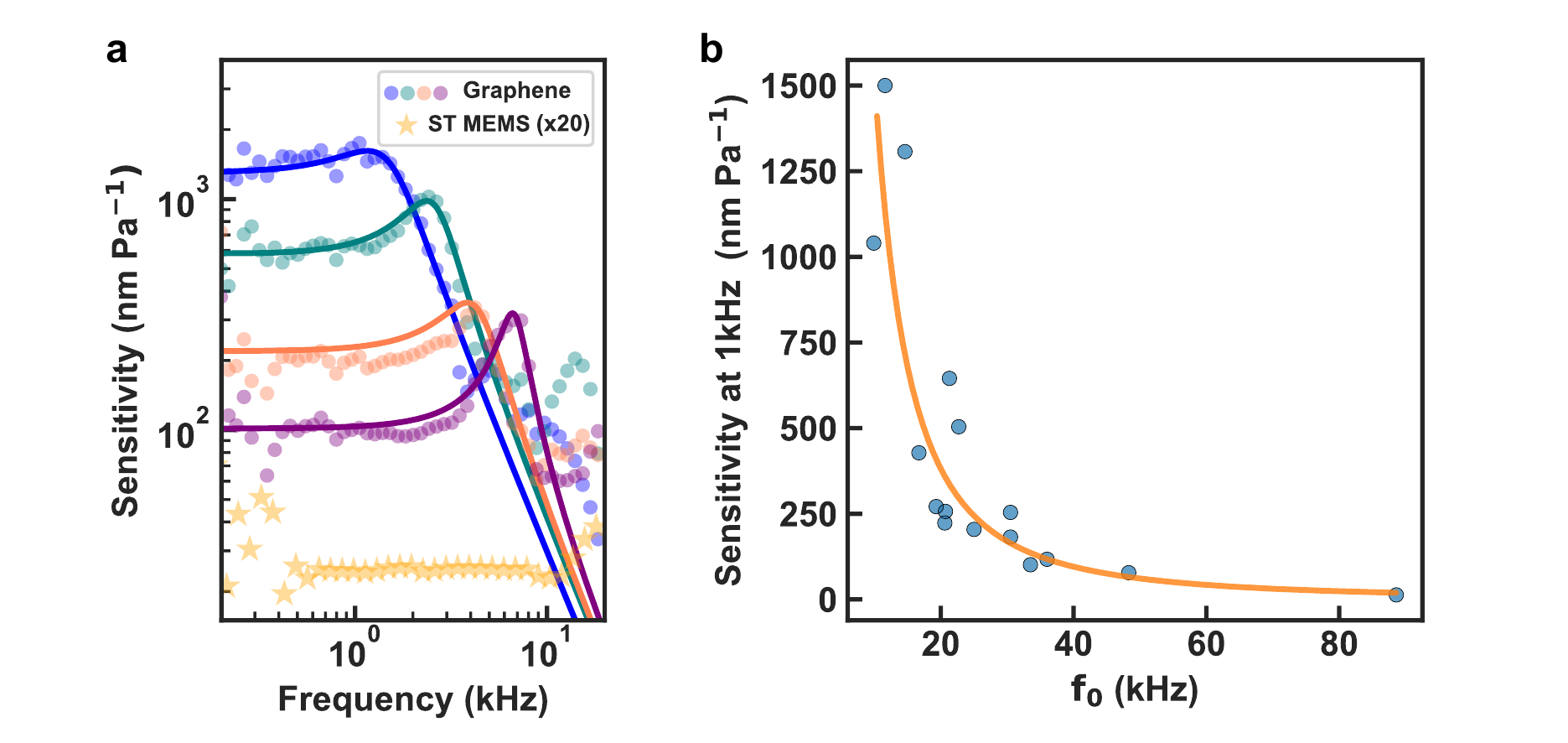}
    \caption{\textbf{Sensitivity in the audible spectrum.} \textbf{a} Audio response spectra of graphene membranes ($d=350 \,\mu$m) and the Si membrane in the ST MEMS microphone. Drawn lines are fits to the data using a harmonic oscillator model. \textbf{b} Acoustic sensitivity of 16 different graphene membranes with $d=350 \,\mu$m at 1 kHz plotted against the fundamental resonant frequency $f_0$ measured in vacuum with a scanning LDV (as discussed in detail in section 4 of the Supplementary Information). Differences in resonance frequency and sensitivity are attributed to variations in pretensions induced by the transfer process.}% \textbf{b} }
    \label{fig:Fig3}
\end{figure}

Following this methodology, we characterize the acoustic response of multilayer graphene membranes of varying diameters, as well as the membrane of a commercial MEMS microphone from ST-Microelectronics (MP23DB01HP), and compare their performance. To avoid confusion with other MEMS devices from literature, the commercial device is referred to as 'ST MEMS microphone' in the rest of the manuscript. In Fig. \ref{fig:Fig3}a, the frequency response of four graphene drums with a diameter $d=$ 350 $\mu$m is shown together with the response of the ST MEMS microphone ($d=950 \,\mu$m). Also, the mechanical sensitivity at 1 kHz ($S_{m,\text{1kHz}}$) of the 37 measured drums is shown in Fig. S3 of the Supplementary Information. This quantity is defined as $S_{m,\text{1kHz}}= \dfrac{\Delta z_{\text{1kHz}}}{\Delta P_{\text{1kHz}}}$  where $\Delta z_{\text{1kHz}}$ is the AC amplitude of the membrane center at 1 kHz and  $\Delta P_{\text{1kHz}}$ is the input sound pressure amplitude at 1 kHz. Even though large differences in sensitivity between graphene membranes are observed,  all graphene membranes exhibit  much higher mechanical sensitivities (up to $\sim 2000$ nm/Pa) than the ST MEMS microphone with $S_{m,\text{1kHz}}\sim 1.3$ nm/Pa.

To analyze these results, the data points in Fig. \ref{fig:Fig3}a were fit (drawn lines) using a harmonic oscillator model, yielding a frequency dependent mechanical sensitivity $S_\text{m}(\omega)$
\begin{equation}
\label{eq:S_m}
    S_\text{m} (\omega) = \frac{R^2}{4 n_0 \sqrt{(1-\omega^2/\omega_0^2)^2 + \omega^2/(\omega_0^2 Q^2)}}\,,
\end{equation}
where $n_0$ is the initial pretension in the membrane, $R$ is the membrane radius, $Q$ is the quality factor and $f_0 = \omega_0/(2 \pi)$ the fundamental resonance frequency corresponding to peaks in the curves in Fig \ref{fig:Fig3}a. The low-frequency response ($\omega \ll \omega_0$), $S_\text{m}(0) =\dfrac{R^2}{4n_0}$, can be calculated using the equation for the static displacement, $z$, of a circular membrane subjected to a uniform pressure load $\Delta P$\cite{makaP}:
\begin{equation}\label{eq:deltap}
    \Delta P = \frac{4n_0}{R^2}z +\frac{8 E t}{3R^4(1-\nu)}z^3\,,
\end{equation}
where $E$ is Young's modulus and $\nu$ is the Poisson's ratio of the material. Thus, in the limit of small $z$, the mechanical sensitivity of the membrane can be expressed as $S_\text{m}(0) \approx \dfrac{z}{\Delta P} =\dfrac{R^2}{4n_0}$. Even though it has a two to three times smaller diameter than the ST MEMS microphone, the mechanical sensitivity of the graphene membrane is extremely large thanks to its low pretension $n_0$.
According to equation (\ref{eq:S_m}), the main parameter determining $S_\text{m}(0)$ is the pretension. Thus, variation in sensitivity observed between the devices in Fig. \ref{fig:Fig3} is likely due to fabrication induced differences in pretension. To check this hypothesis, we consider the equation for the fundamental resonance frequency, $f_0$, of a circular membrane and its relation to the mechanical sensitivity $S_\text{m}(0)$:
\begin{equation}
    f_0 = \frac{2.405}{2\pi R}\sqrt{\frac{n_0}{\rho_\text{eff}t}}= \frac{2.405}{4\pi}\sqrt{\frac{1}{ S_\text{m}(0) \rho_\text{eff}t}}\,,
    \label{eq:Svsf}
\end{equation}
where $\rho_\text{eff}$ is the effective density of the membrane, which can be affected by air loading effects (see section 4 of the Supplementary Information). Since 1 kHz is below the resonance frequency of the membrane, $S_{\text{m},\text{1kHz}} \approx S_\text{m}(0)$. Therefore, we expect to observe the following proportionality relation between mechanical sensitivity and resonance frequency: $S_\text{m} \propto f_0^{-2}$.

To remove the influence of air loading effects on the resonance frequency, we also measure the membranes' resonance frequency in vacuum using a scanning LDV. To determine the membrane's resonance frequency in vacuum, a scanning laser Doppler vibrometer is used (MSA400 Micro System Analyzer). The sample is placed inside a vacuum chamber ($\sim 10^{-3}$ mbar) equipped with a piezo shaker to actuate the membrane. The displacement is measured over a user-defined grid of points distributed over the surface of the membrane. Thus the membrane mode shape can be reconstructed to identify the first resonance mode (see section 4 of the Supplementary Information for more details). 

In Fig. \ref{fig:Fig3}b we plot the sensitivity at 1 kHz against  $f_0$ measured using a scanning LDV in vacuum. The data in Fig. \ref{fig:Fig3}b follows the theoretically expected relation $S_\text{m} \propto f_0^{-2}$, showing that the experimental differences in sensitivity observed in Fig. \ref{fig:Fig3}b can indeed be well accounted for by variations in $n_0$. Figure S3b in the Supplementary Information, shows correlations between mechanical sensitivity and resonance frequencies measured in air from data like in Fig. \ref{fig:Fig3}a.  Variations in pretension can be caused by forces on the graphene during the transfer process, and might also be induced by wrinkles in the membranes (see Fig. S5 in the Supplementary Information).

\begin{figure}[H]
\includegraphics[width = 0.8\linewidth]{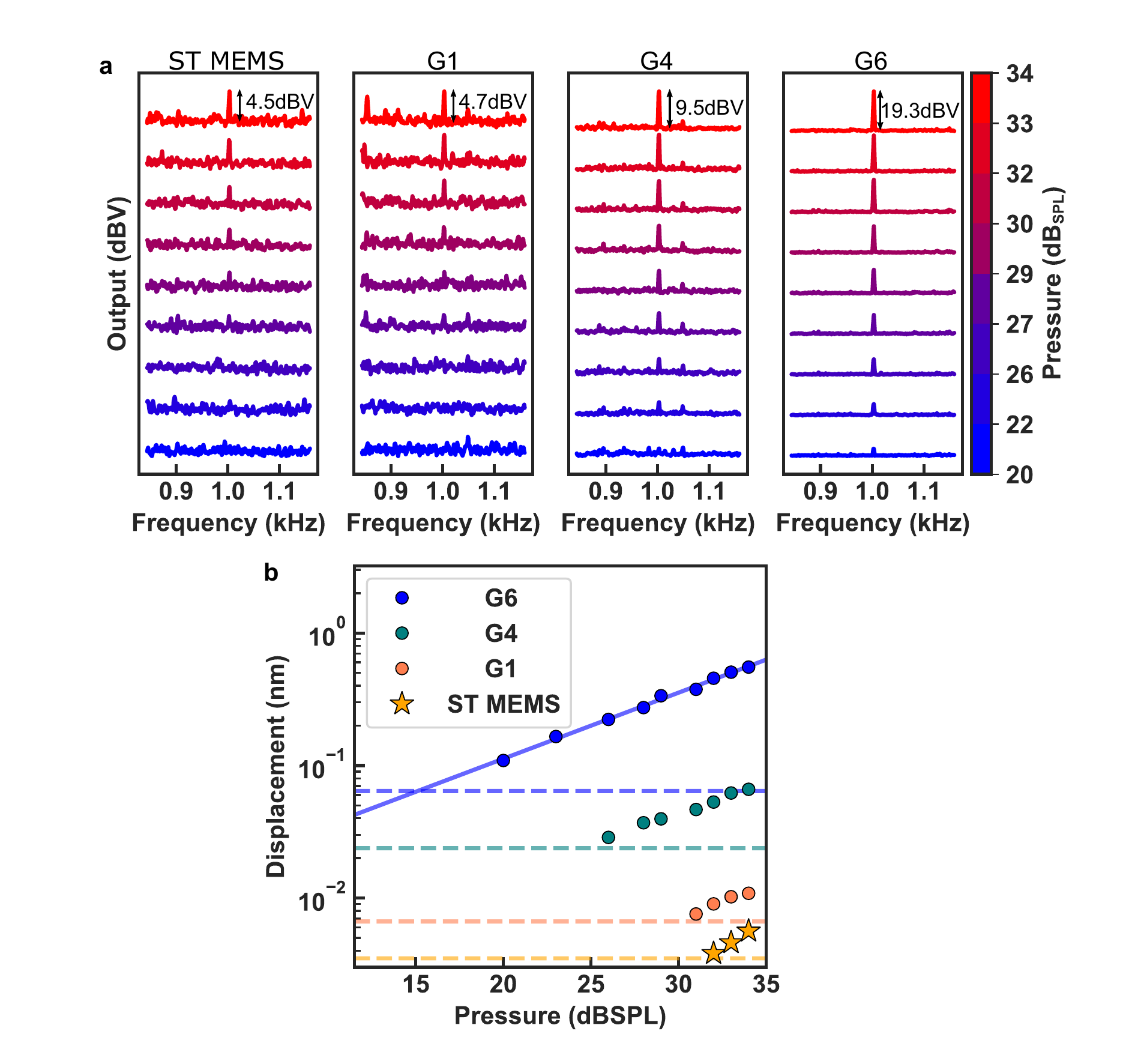}
\caption{\textbf{Minimum detectable SPL.} \textbf{a} Waterfall plot of the displacement amplitudes for three graphene drums and the ST MEMS microphone in response to a 1 kHz tone at low input sound pressure levels ($20-34$ dB$_\mathrm{SPL}$ indicated by color scale). The most sensitive graphene membrane G6 detects sounds down to 20 dB$_\mathrm{SPL}$ with SNR = 5. \textbf{b} Extracted peak amplitudes at 1kHz from Fig. \ref{fig:Fig4}a of the four samples. The dashed lines show the average noise level (NL) of the corresponding samples. Peaks with amplitudes smaller than 1.1$\times$NL were removed from the plot. A linear fit through the points from device G6 is also shown to determine the minimum detectable SPL by extrapolation (LOD $\sim$ 15 dB$_\mathrm{SPL}$).} 
\label{fig:Fig4}
\end{figure}

\begin{figure}[H]
\includegraphics[width = 0.8\linewidth]{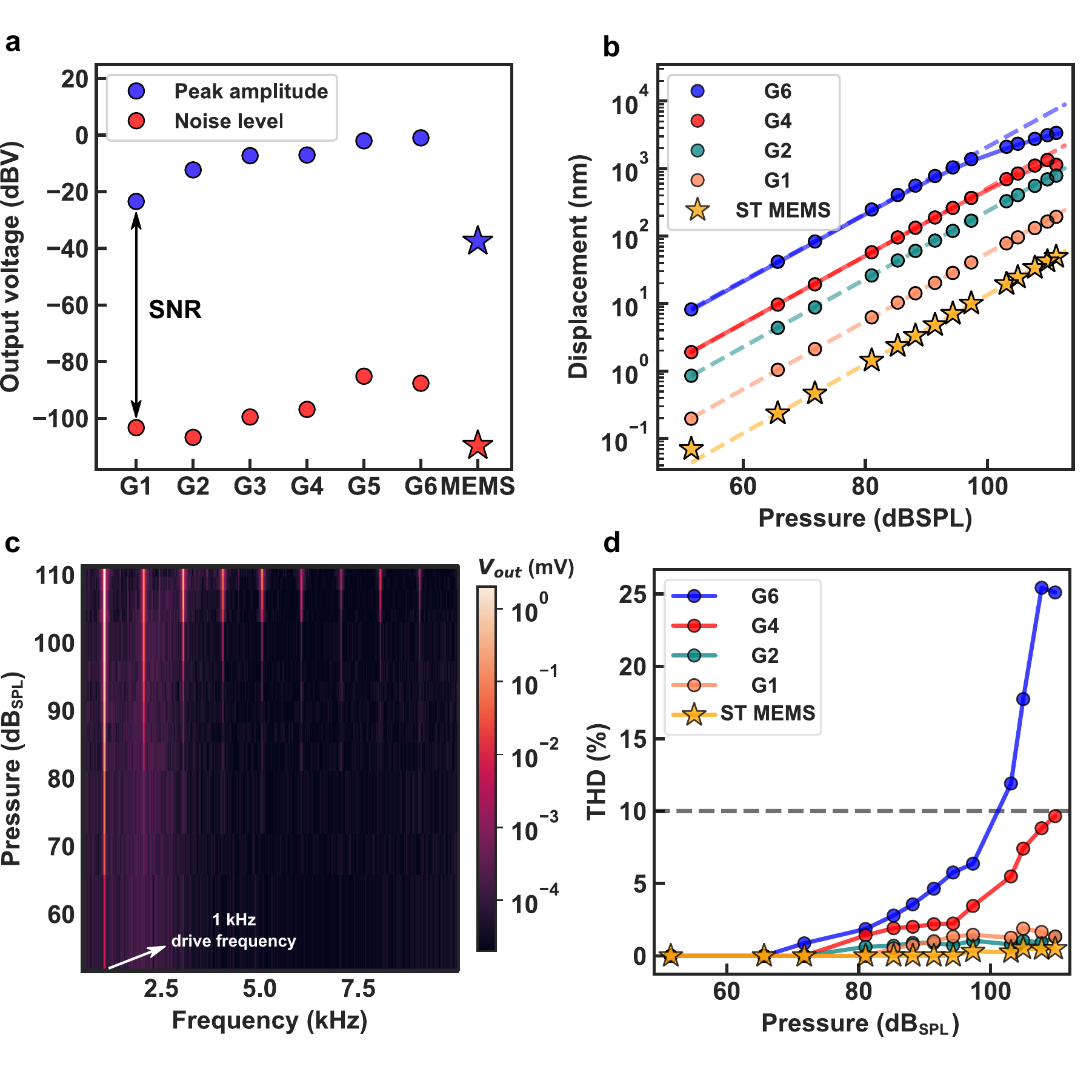}
\caption{\textbf{Signal-to-noise ratio and harmonic distortion.} \textbf{a} Comparison between displacement signal from graphene membranes (circles) and from the ST MEMS microphone (stars) in response to a 1 kHz tone at 1 Pa of rms SPL (= 94 dB$_\mathrm{SPL}$). The blue markers indicate the peak amplitude while the red markers indicate the noise floor of the spectra (like in Fig. \ref{fig:Fig4}a). Average SNR in graphene is 88 dB, 16 dB higher compared to that of the ST MEMS microphone. \textbf{b} Displacement amplitude at 1kHz vs. SPL for several graphene drums and the ST MEMS microphone, extracted from spectra like in Fig. \ref{fig:Fig4}a. \textbf{c} Displacement spectrum of device G6 as a function of SPL of a 1 kHz tone from the speaker. \textbf{d} Total harmonic distortion (THD) versus SPL for the samples in \ref{fig:Fig5}b. The dashed line at $\text{THD}=10$\% marks the acoustic overload point (AOP).}
\label{fig:Fig5}
\end{figure}

A high sensitivity does not automatically guarantee that a microphone can detect weak sounds, because its limit of detection (LOD) also depends on the noise level. To determine the LOD, we measured the membrane displacement at  1 kHz for different driving amplitudes to investigate the minimum detectable sound pressure level (SPL).
Figure \ref{fig:Fig4}a shows the displacement signal from the vibrometer in response to a 1 kHz tone at low SPL ($<35$ dB$_\mathrm{SPL}$) for the ST MEMS microphone and three graphene membranes with different mechanical sensitivity (labelled as G1, G4, G6) and $d=350\,\mu$m. The vibration amplitude at 1 kHz as a function of the input SPL, as obtained from the peak heights in Fig. \ref{fig:Fig4}a using the vibrometer responsivity of 200 nm/V, is shown in Fig. \ref{fig:Fig4}b for the four devices. The measured average noise level for each sample is depicted with a dashed line of the corresponding color. For SPL $>30$ dB$_\mathrm{SPL}$, the response peak at 1 kHz is visible in all samples with varying amplitudes depending on the sample's mechanical sensitivity. When decreasing the input SPL, the 1 kHz peak becomes comparable to the noise level at $\sim 30-32$ dB$_\mathrm{SPL}$ for the ST MEMS microphone and G1, while for G4 and G6 the extrapolated signal stays above the noise level down to 25 and 15 dB$_\mathrm{SPL}$ respectively, which is significantly lower than the lowest SPL of 70 dB$_\mathrm{SPL}$ at which graphene membranes were tested in literature\cite{wood_design_2019} up to now. 
The extrapolated LOD of $\sim 15$ dB$_\mathrm{SPL}$ of device G6 is even lower than the specified LOD of the reference microphone of 24 dB$_\mathrm{SPL}$ used to measure the input SPL.
Figure \ref{fig:Fig4}b shows that the noise level increases when the sensitivity increases due to a smaller stiffness $k$, because the thermomechanical noise induces a mean displacement given by:\cite{gabrielson_mechanical-thermal_1993} $<x^2> = \dfrac{k_B T}{k}$. The thermomechanically induced displacement power spectral density below resonance $S_{xx,n}=\dfrac{4 k_B T}{k Q \omega_0}$ can be calculated by extracting $k$ from a linear fit to Figure \ref{fig:Fig5}b and $Q$ and $\omega_0$ from a harmonic oscillator fit to the resonances in air, leading to a theoretical value of the thermomechanical noise displacement density $\sqrt{S_{xx,n}}$ of $\sim$ 8, 1.8, 0.2 pm$/\sqrt{\text{Hz}}$ for sample G6, G4, G1 respectively. The noise level measured is $\sim$ 47, 17, 5 and 3 pm$/\sqrt{\text{Hz}}$ for sample G6, G4, G1  and the ST MEMS respectively, showing that the displacement noise in the membranes is near, but not at the theoretical limit.
To determine the microphone performance at high sound pressure levels, similar measurements were performed at high SPL to study the dynamic range, the distortion and nonlinearity of the response. In Fig. \ref{fig:Fig5}a we show the response amplitude as well as the average noise level of some membranes to a 1kHz tone of 1 Pa (= 94 dB$_\mathrm{SPL}$) to compare their signal-to-noise ratio SNR$_{1 \text{Pa}, 1 \text{kHz}}=x_{1 \text{Pa}, 1 \text{kHz}}/\sqrt{S_{xx,n}}$ to that of the ST MEMS microphone. On average, the noise level of the graphene membranes is higher compared to that of the ST MEMS microphone. However, due to their higher sensitivity at 94 dB$_\mathrm{SPL}$, the SNR (difference between blue and red data points in Fig. \ref{fig:Fig5}a) of the graphene microphones ranges from 80-95 dB, which is significantly larger than that of the ST MEMS microphone, which is 72 dB.

In Fig. \ref{fig:Fig5}b, we show the peak amplitude of the displacement signal in response to a 1 kHz tone between 50 dB$_\mathrm{SPL}$ and 110 dB$_\mathrm{SPL}$. Louder acoustic signals were not used due to large distortion and clipping occurring in the speaker. All graphene samples exhibit a higher response than the ST MEMS microphone, but at a high SPL the most sensitive samples deviate from linear behaviour. This is to be expected as the linear approximation of equation (\ref{eq:deltap}) holds in the limit of small displacements. Therefore, while sample G6 was the best at detecting low sound levels down to 20 dB$_\mathrm{SPL}$, its performance at high SPL gets worse due to non-linear effects limiting its dynamic range. The non-linear response of G6 and G4 was fitted to equation (\ref{eq:deltap}) with $t=8$ nm, $2R = 350\,\mu$m and $\nu =0.26$ yielding a pretension $n_0 \sim$ 7 mN/m and 33 mN/m and a Young's modulus of $E \sim$ 5 GPa and 30 GPa for G6 and G4 respectively. The extracted Young's modulus is much lower than that of pristine graphene. This reduction could be due to defects that originate from  graphene growth in the form of small holes of $\sim$ 50 nm, as observed by SEM inspection of the samples (see Fig. S6 in supplementary information). In addition, transfer-induced wrinkles and slack in the membrane can further decrease the Young's modulus \cite{li_deformation_2015}.

To analyze the observed distortion, the spectrum of sample G6 in response to a 1 kHz tone with varying SPL at 1 kHz is shown in Fig. \ref{fig:Fig5}c. At higher SPL, harmonics of the driving frequency are visible in the spectrum. In order to measure the distortion level and maximum detectable SPL of the samples under study, we calculated the THD from the first five harmonics as\cite{inve}:
\begin{equation}
    \text{THD} = \frac{\sqrt{\sum_{i>0} V_i^2 }}{V_0}\,,
\end{equation}
where $V_i$ is the rms voltage output at the $i$-th harmonic ($i>0$) and $V_0$ is the output at the driving frequency. The maximum SPL that a microphone can handle is determined by the acoustic overload point (AOP), which is usually defined as the pressure level at which the THD reaches 10\% \cite{inve}. The dynamic range is then given by the difference in dB between the AOP and minimum detectable SPL, which for MEMS devices is usually determined by the microphone's noise level. In the calculation of the THD of the membrane's displacement, contributions from  harmonics generated by the speaker were subtracted. The extracted THD as a function of the input SPL is shown in Fig. \ref{fig:Fig5}d. Samples G6 and G4 reach the acoustic overload point at around 98 dB$_\mathrm{SPL}$ and 110 dB$_\mathrm{SPL}$ respectively, setting their dynamic range to $\sim$ 83 dB and 87 dB. The other samples are well below the AOP at the maximum SPL applied so that their dynamic range is $>$ 85 dB. The measured THD at 110 dB$_\mathrm{SPL}$ of the ST MEMS microphone is $\sim 0.5\%$, in close agreement with the reported THD in its datasheet\cite{STM}.

In order to theoretically estimate the expected value of the THD, we assume periodic motion of the membrane $z(t) = z_0\sin\omega t$, which when substituted in equation (\ref{eq:deltap}) yields
\begin{equation}
    \Delta P(t) = \left(\alpha z_0 + \frac{3}{4}\beta z_0^3\right)\sin\omega t - \frac{\beta}{4}z_0^3\sin 3\omega t\,,
\end{equation}
where $\alpha = \dfrac{4n_0}{R^2}$ and $\beta = \dfrac{8Et}{3R^4(1-\nu)}$. Using values of $n_0$ and $E$ from the fit to the non-linear response curves in Fig \ref{fig:Fig5}b, we calculate the expected distortion just from the third harmonic as: $\text{THD}_{3\text{rd}} = \dfrac{\beta z_0^2}{4\alpha + \beta z_0^2}\,$. The resulting $\text{THD}_{3\text{rd}}$ for samples G6 and G4 at the maximum applied SPL of 110 dB$_\mathrm{SPL}$ is 21.5\% and 7.5\% respectively. This is consistent with the measured values of 25\% and 9.6\% corresponding to THD from the first 5 harmonics, since the third harmonic is expected from equation (2) to have the largest contribution to the THD.

\section{Discussion}
We have established that graphene membranes can offer very high acoustical sensitivities that are up to two orders of magnitude higher than that of MEMS microphones.  In commercial applications, a microphone’s signal is usually detected electrically via capacitive readout.
In the capacitive configuration, the microphone's ability to pick up sound is described by its open circuit sensitivity, $S$, namely the ratio of the microphone’s open circuit output voltage over the driving SPL in Pa. The open circuit sensitivity is given by the product of the microphone's electrical $S_\text{e}$ and mechanical $S_\text{m}$ sensitivities as \cite{scheeper_review_1994}:
\begin{equation}
\label{eq:SeSm}
    S = S_\text{e}\, S_\text{m} = \frac{V_\text{b}}{g_0}S_\text{m} \,,
\end{equation}
where $V_\text{b}$ is the bias voltage and $g_0$ is the equilibrium gap distance between the capacitor plates. Using this formula we can further compare our membranes with MEMS devices reported in literature\cite{zawawi_review_2020} by extracting their mechanical sensitivity $S_m$ from published values of $S$, $V_\text{b}$ and $g_0$ using equation (\ref{eq:SeSm}). To have a more direct comparison between devices of different geometries/dimensions, we divide the mechanical sensitivity by the area of the membrane which yields the membrane's mechanical compliance, $C_\text{m}$. 
\begin{figure}[H]
    \centering
    \includegraphics[width = 0.75\linewidth]{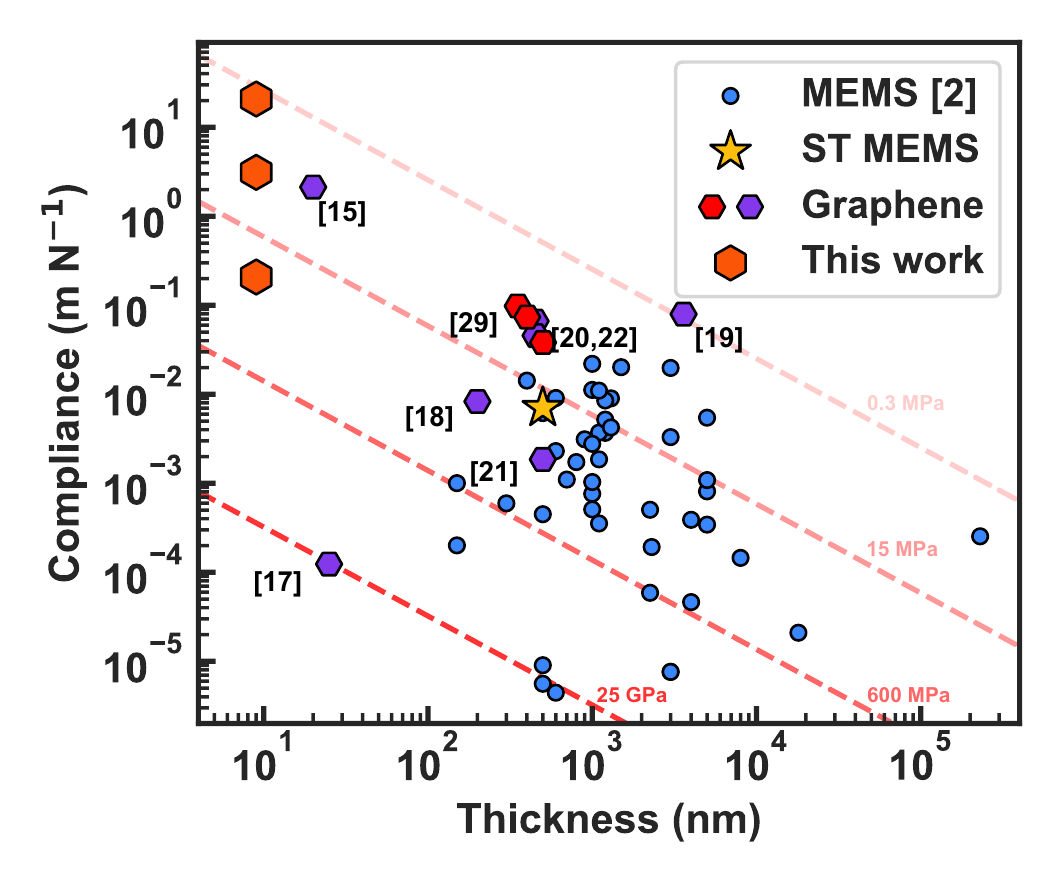}
    \caption{\textbf{Mechanical compliance of literature devices.} Scatter plot of mechanical compliance vs. membrane's thickness for MEMS microphones from literature\cite{zawawi_review_2020} (blue circles), the ST MEMS microphone (yellow star), graphene microphone literature (purple and red hexagons for membranes with and without a backplate respectively) and for three graphene membranes in this work (orange hexagon), including membranes with lowest and highest compliance measured. The dashed red lines indicates lines with constant stress, with lower stress corresponding to lines with lower opacity. The relevant reference numbers for the graphene microphones are indicated in the graph near the data points.}
    \label{fig:Fig6}
\end{figure}
Figure \ref{fig:Fig6} shows the mechanical compliance as a function of membrane thickness for MEMS devices reported in [\citen{zawawi_review_2020}] (blue circles), for graphene-based microphones (red and purple hexagons), and for the multi-layer graphene (MLG) membranes presented here (orange hexagon). The data points shown for the MLG membranes of our work include the membranes with highest and lowest compliance to highlight the range of measured values compared to previous reports in literature. The compliance of graphene-based membranes not listed in [\citen{zawawi_review_2020}] is estimated from the reported membrane pretension using $C_\text{m} = \dfrac{1}{k}$, where the membrane's stiffness $k = 4\pi n_0$ in the case of pressure deformation \cite{davidovikj_nonlinear_2017}. Moreover, since $n_0 = \sigma t$ where $\sigma$ is the pre-stress of the membrane, we can identify lines of constant stress in the $C_\text{m}$ vs. $t$ plane. These are highlighted as dashed red lines in Fig. \ref{fig:Fig6} with lower stress corresponding to lower opacity of the line. The high mechanical sensitivity of our membranes can thus be attributed to a combination of low stress and small thickness. The high stress reported on the $t = 25$ nm membrane in Ref. [\citen{todorovic_multilayer_2015}] results from the large polarization voltage of 200 V used to readout the acoustic signal, which led to an estimated pretension of 640 N/m. 

We note that most microphone works in literature deal with membranes with a backplate for capacitive readout. Therefore, the lower compliance in these devices (at atmospheric pressure) can be partly explained by the effect of squeeze film damping. The graphene membranes under study do not have such a backplate, because we first wanted to determine the intrinsic properties of the graphene membranes themselves. Realizing efficient electrical microphone readout, e.g. via a perforated capacitive backplate, while maintaining this high sensitivity and compliance is another challenge that is outside the scope of this work.

Although lowering the tension and stiffness of graphene membranes helps to improve their acoustic sensitivity, a drawback is that it reduces the fundamental resonance frequency, thus limiting the microphone bandwidth (Fig. \ref{fig:Fig3}b). Since the mass of the graphene membranes is extremely low, and their aspect-ratio very high ($d/t \sim$ 46000), the mass of the air that moves along with the membrane is substantial, increasing the effective membrane mass\cite{al-mashaal_dynamic_2017, she_effect_2016} $m_\text{eff}$. This mass increase further reduces the resonance frequency and bandwidth. Initial experiments showed a 5 to 9-fold decrease of the resonant frequency from vacuum $\sim 10^{-4}$ mbar to atmospheric conditions (see Fig. S4 in the Supplementary Information). Further pressure-dependent measurements are needed to understand better this air mass loading effect as well as the importance of squeeze-film damping on future devices with backplate for capacitive readout.

In general, it is desirable to have the resonance frequency of the membrane above the audible range ($>$ 20 kHz). The membranes in this work, like in several other graphene microphone publications\cite{xu_realization_2021, woo_realization_2017,zhou_graphene_2015}, do not satisfy this condition. However, it is important to note that depending on the target application, a bandwidth of 6-10 kHz can be sufficient\cite{fueldner_microphones_2020, woo_realization_2017}, therefore the reported low resonance frequency ($<$10 kHz) of graphene membranes is not necessarily limiting for their performance. Nevertheless, this problem could be compensated for in next generation devices by increasing their pretension $n_0$ or reducing the membrane radius $R$, while keeping competitive mechanical sensitivity. For example, as shown in Ref. [\citen{RobertoPaper}] graphene membranes with diameters of 85-150 $\mu$m exhibit resonance frequencies in vacuum of 250-320 kHz and mechanical sensitivities still comparable to a MEMS membrane with diameter of 950 $\mu$m. For a fairer comparison, one can correct the obtained compliances in Fig. \ref{fig:Fig6} by a factor $(20kHz/f_0)^2$. Even after such a correction, the compliances obtained by the graphene membranes this work are higher than most literature values as shown in Fig. S4 in the Supplementary Information. 

%Nevertheless, thanks to the high mechanical sensitivity of graphene membranes, it is possible to further reduce their size to achieve higher resonance frequency and bandwidth while keeping good acoustic performance  The effect of added air mass would also be reduced in smaller membranes (Eq. 1 in Supplementary Information) as well as squeeze film damping in closed cavity resonators with capacitive readout\cite{BAO20073}.

A main challenge in using graphene as a microphone is linked to the lack of control over its mechanical properties during the transfer process, which limits the reproducibility of the membrane's performance as shown in Fig. \ref{fig:Fig3} and S3. For microphone applications, large sheets of suspended CVD graphene are needed and thus a transfer step to the target substrate has been unavoidable in all previous studies. In addition to the poor uniformity and control of strain, the transfer process can degrade the quality of the graphene by introducing contamination, cracks and wrinkles, unwanted for practical application and large-scale production\cite{wagner_graphene_2016}. In a recent study \cite{RobertoPaper}, wafer-scale fabrication of multilayer graphene membranes was achieved using a transfer-free method, by which the graphene is grown and released directly on the target substrate. This novel method could prove beneficial in terms of uniformity and scalability in fabrication of graphene-based microphones and sensors. 

Finally, the most sensitive membranes are found to be more influenced by non-linear effects at high SPL, and exhibit higher distortion and reduced dynamic range. Graphene membranes cannot yet reach commercial values of THD, acoustic overload point (AOP=140 dB$_\mathrm{SPL}$) and dynamic range (105 dB$_\mathrm{SPL}$\cite{STM}), also because they do not feature a double-backplate configuration for differential readout which greatly reduces the THD and increases sensitivity\cite{fueldner_microphones_2020}. The trade-off between sensitivity and dynamic range could be further optimized by better control over the membrane's stiffness. 

\section{Conclusions}
In this work, evidence is provided that, with proper design, graphene-based devices have the potential to outperform existing microphones. %Using laser Doppler vibrometry, we have measured the frequency response over the audible spectrum (200 Hz to 20 kHz) as well as the membrane displacement as a function of input sound pressure level (SPL) at 1 kHz of several graphene membranes with diameters ranging from 350 to 570 $\mu$m and thickness of around 8 nm, and compared their performance to a commercial MEMS microphone with a diameter of 950 $\mu$m. 
In terms of mechanical sensitivity and SNR, graphene is superior to commercial Si-based membranes and MEMS devices from literature by a large margin, yielding an improvement of more than 2 orders of magnitude in sensitivity. In addition we show that the detection limit of graphene membranes is as low as 15 dB$_\mathrm{SPL}$ for membranes with a diameter of only $350 \,\mu$m. On the other hand, due to the low stiffness and the large contribution of air loading, the membrane's bandwidth is found to be limited at $<10$kHz on most samples.  However, we show that even when taking this factor into account, the membranes in this work are still more performant than commercial devices. We propose that given the high sensitivity of graphene one can further reduce the sensor's dimensions, increasing its resonance frequency, bandwidth and mechanical strength. Smaller sized membranes would also facilitate the implementation of arrays of membranes in parallel to increase SNR\cite{fueldner_microphones_2020} or directionality and reduce effects of added air as well as squeeze film damping in closed cavity resonators with capacitive readout\cite{BAO20073}. Therefore, we believe that graphene can indeed improve current microphone devices and that the main disadvantage and barrier to real applications lies in the lack of a more controllable fabrication method to suspend graphene membranes.

\begin{acknowledgement}
The authors acknowledge funding from European Union’s Horizon 2020 research and innovation program under Grant Agreement No, 881603 (Graphene Flagship) and No. 883272 (BorderUAS) and from the Ministry of Science, Education, and Technological Development of the Republic of Serbia through grant no. 451-03-68/2022-14/200026.
\end{acknowledgement}

\section*{Author contributions}
G.B., H.S.J.v.d.Z. and P.G.S. conceived the experiments. G.B. performed the experiments. R.P. and K.C fabricated and inspected the samples. G.B., G.J.V, H.L, and P.G.S. analyzed and modeled the experimental data. G.B wrote the manuscript with contributions from all the authors. H.S.J.v.d.Z., M.S., D.T, G.J.V., S.V. and P.G.S. revised the manuscript. H.S.J.v.d.Z., M.S., S.V. and P.G.S. supervised the project. 

\section*{Data availability}
The data that support the findings of this study are available from the corresponding authors upon request.

\section*{Competing interests}
The authors declare no competing interests.

\section*{Additional information}
\textbf{Supporting information}
\begin{itemize}
    \item Figure S1: Raman spectrum of graphene samples.
    \item Figure S2: AFM characterization of graphene samples.
    \item Figure S3: Mechanical sensitivities at 1 kHz of all measured samples.
    \item Figure S4: Measurements of resonance frequency in vacuum and comparison with resonance in air.
    \item Figure S5: Effects of wrinkles on membrane's mode shape.
    \item Figure S6: SEM pictures of graphene membranes.
    \item Figure S7: Mechanical compliance corrected by resonance frequency
    \item Movie V1: LynceeTec mode shape imaging of graphene membrane.
    \item Movie V2: LynceeTec mode shape imaging of wrinkled graphene membrane.
    \item Audio A1: music track (Ode to joy) recorded on optically on multilayer graphene membrane.
\end{itemize}

\noindent\textbf{Correspondence} and requests for materials should be addressed to G.B. or P.G.S.

\newpage

\renewcommand{\figurename}{Figure S}
\setcounter{figure}{0}    
\section{S1. Performance parameters}
Here we list the definition of relevant microphone performance parameters addressed in the main manuscript. Following common conventions for microphone specifications, the reference input frequency and reference pressure level are taken to be 1kHz and 1 Pa = 94 dB$_\mathrm{SPL}$ respectively.
\begin{itemize}
    \item  \textbf{Sensitivity}: ratio between electrical output and input sound pressure, usually expressed in mV/Pa for capacitive microphones. This overall sensitivity is a combination of \textit{electrical sensitivity}, which depends on the readout-circuit and amplification, and \textit{mechanical sensitivity}\cite{zawawi_review_2020}. In this work, we address the mechanical sensitivity, given by the ratio between membrane's displacement and input pressure which strongly depends on the material properties and membranes dimensions.
    \item \textbf{Signal-to-noise ratio (SNR)}: ratio between output in response from a reference signal (1kHz at 1 Pa) and noise level of the microphone.
    \item \textbf{Dynamic range}: difference between the maximum and minimum sound pressure level that the microphone can handle. The maximum detectable sound is determined by the amount of nonlinear distortion in the microphone response.
    \item \textbf{Total harmonic distortion (THD)}: measures the level of distortion at the output and it is defined as the ratio between the sum of the powers of the harmonics and the fundamental tone. Maximum detectable sound is determined by the acoustic overload point (AOP) corresponding to THD = 10 $\%$.
\end{itemize}

\section{S2. Raman and AFM}
\begin{figure}[H]
    \centering
    \includegraphics[width = 0.7\linewidth]{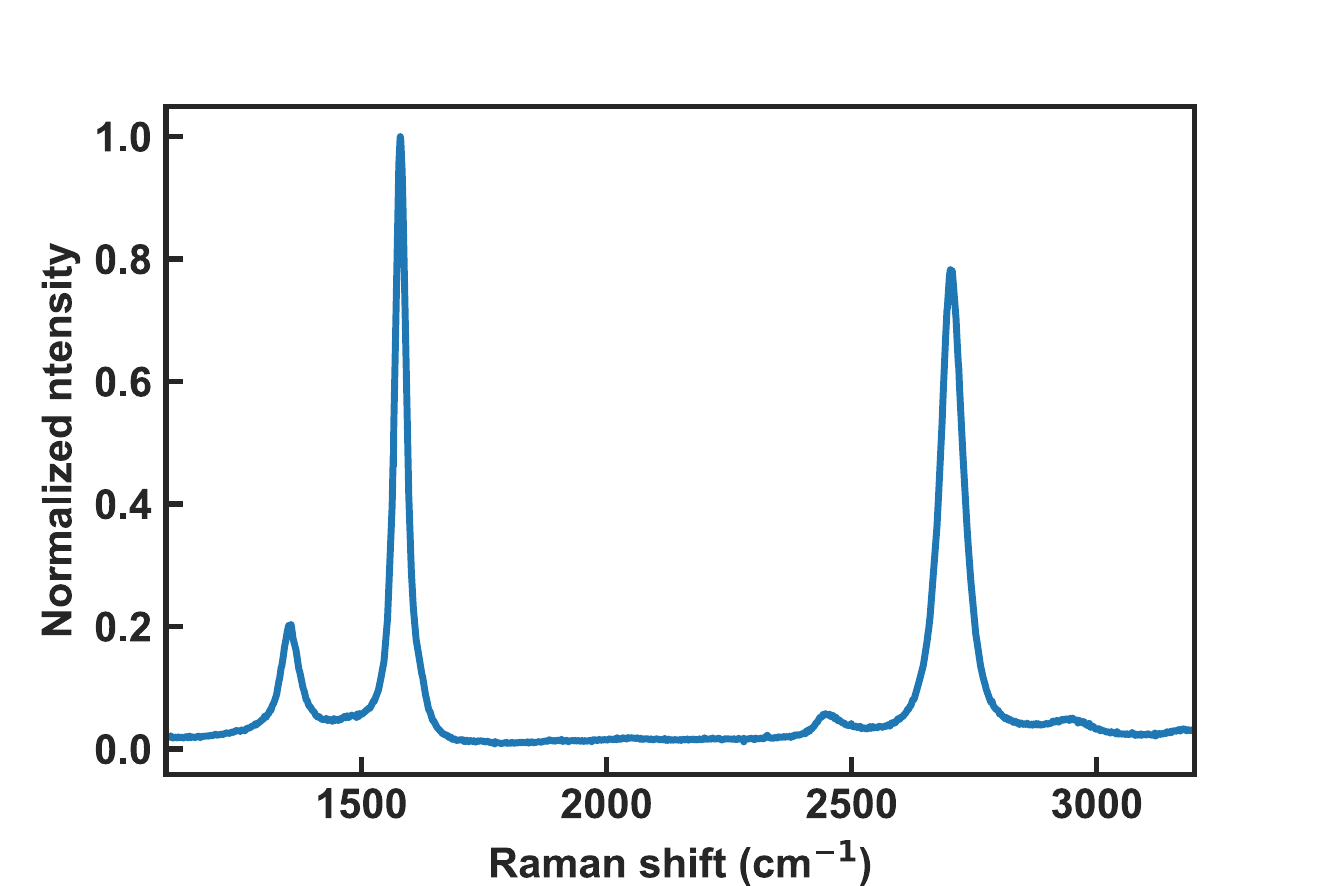}
    \caption{\textbf{Raman spectrum.} Raman spectrum of the CVD multilayer graphene.}
    \label{fig:raman}
\end{figure}

\begin{figure}[H]
    \centering
    \includegraphics[width = \linewidth]{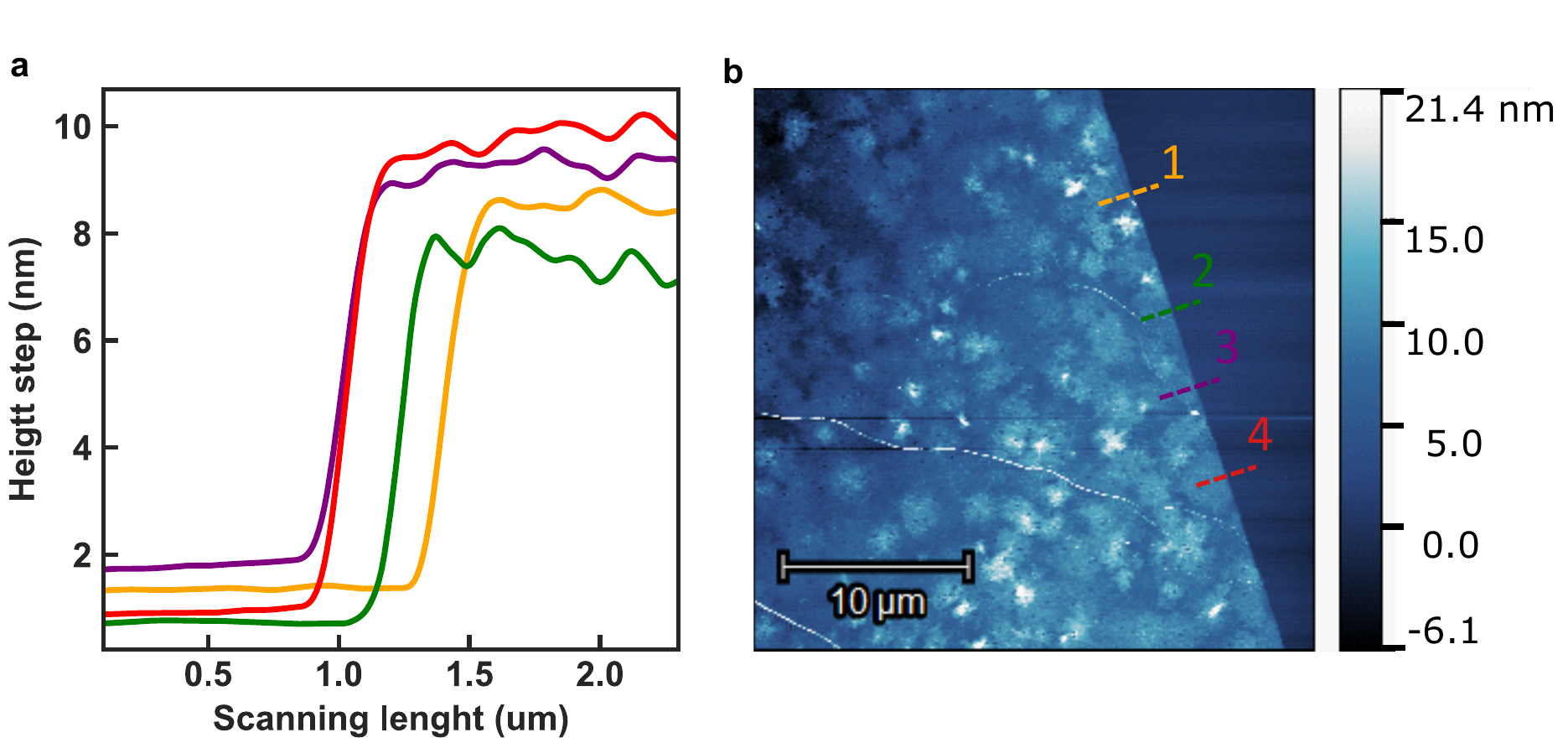}
    \caption{\textbf{Thickness measurements.}(a) Traces of AFM scans at the edge of the graphene sheet after transfer on the Si/SiO$_2$ substrate. (b) AFM map of the scanned area. The coloured lines highlight the regions where the step in Fig. \ref{fig:afm}a is measured.}
    \label{fig:afm}
\end{figure}

The crystallinity of the graphene is inspected by Raman spectroscopy using a Horiba HR800 spectrometer equipped with a 514.5 nm Ar+ laser maintained at 5 mW to limit possible degradation of the material. The objective used is 100x with a numerical aperture of 0.9, giving a spot size of about 696 nm. The Raman spectrum, shown in Fig. \ref{fig:raman}, is normalized to the amplitude of the G band. A total of twelve measurement points in the supported region (graphene on SiO2) show the three typical Raman bands that identify the graphene. The G-band and 2D-band are centered at $\sim$ 1579.8 cm$^{-1}$ and 2703.1 cm$^{-1}$ with a $\text{I}_{\text{2D}}/\text{I}_\text{G} < 1$ that confirms its multi-layer nature \cite{malard_raman_2009, pimenta_studying_2007} . The rise of the D-band at 1354.9 cm$^{-1}$, with $\text{I}_{\text{D}}/\text{I}_\text{G} < 0.2$, shows the presence of defects in the material due to a locally distorted graphene lattice, like edges, vacancies, Stone-Wales defects, and wrinkles. 
We measured the graphene thickness with an atomic force microscope (AFM) from Cypher Asylum Research in semi-contact mode. Multiple profile scans of at the edge of the graphene sheet yield a thickness of $\sim 7.7 \pm 0.8$ nm and are shown in Fig. \ref{fig:afm}.
\section{S3. Additional data on mechanical sensitivity}
\begin{figure}[H]
    \centering
    \includegraphics[width = \linewidth]{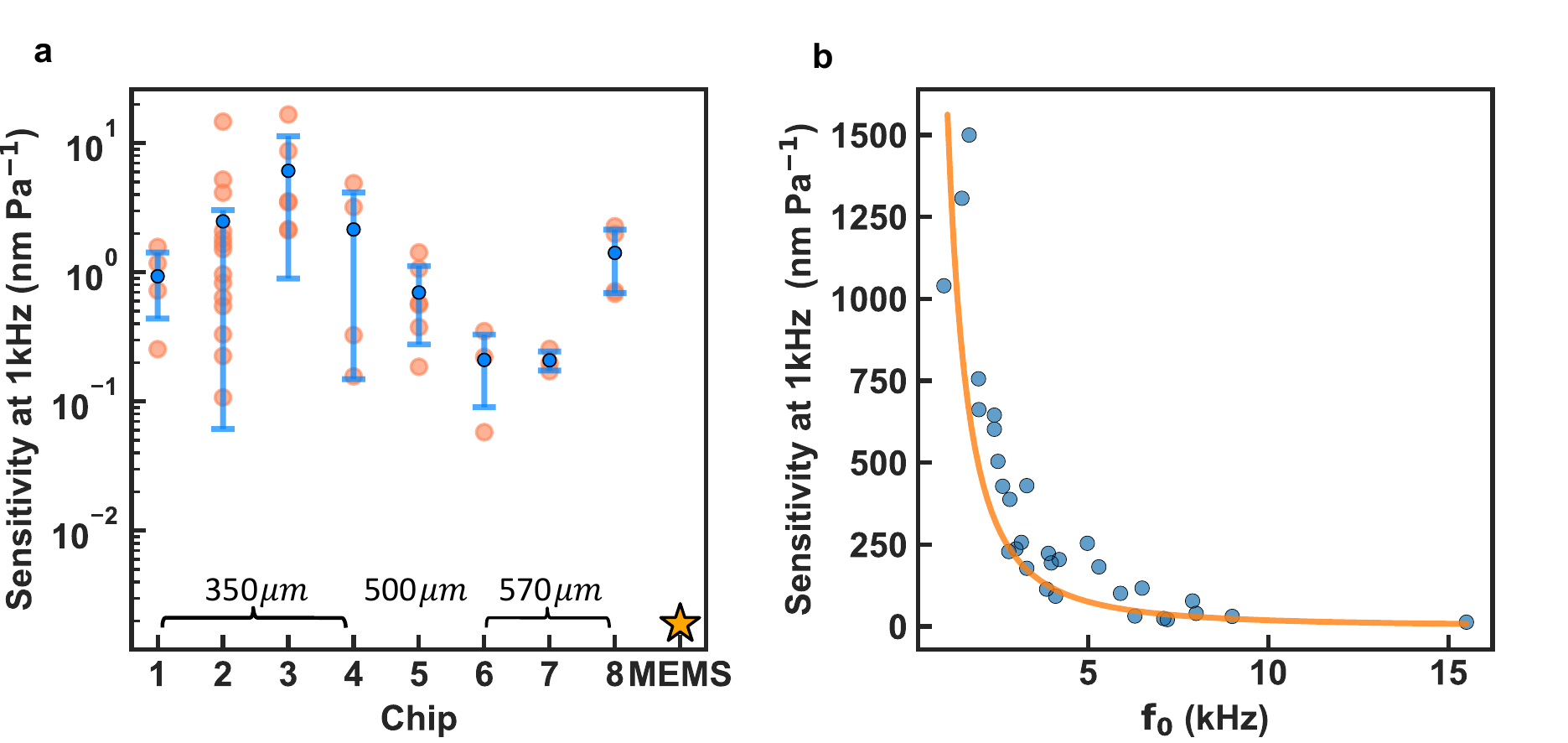}
    \caption{\textbf{Mechanical sensitivity.} \textbf{a} Mechanical sensitivities at 1 kHz of 37 graphene membranes on different chips. The performance varies significantly from sample to sample but all graphene membranes exhibit a higher sensitivity than the reference MEMS microphone. \textbf{b} Acoustic sensitivity of graphene membranes with $d=350 \,\mu$m at 1 kHz plotted against the fundamental resonant frequency $f_0$ measured in air from data like in Fig. 3a of the main text.}
    \label{fig:sens}
\end{figure}

Figure \ref{fig:sens} shows the mechanical sensitivity at 1 kHz measured on all 37 graphene drums as well as the correlation between sensitivity and resonant frequency measured in atmospheric conditions.

\section{S4. Resonance frequency and air damping}
\begin{figure}[H]
    \centering
    \includegraphics[width = \linewidth]{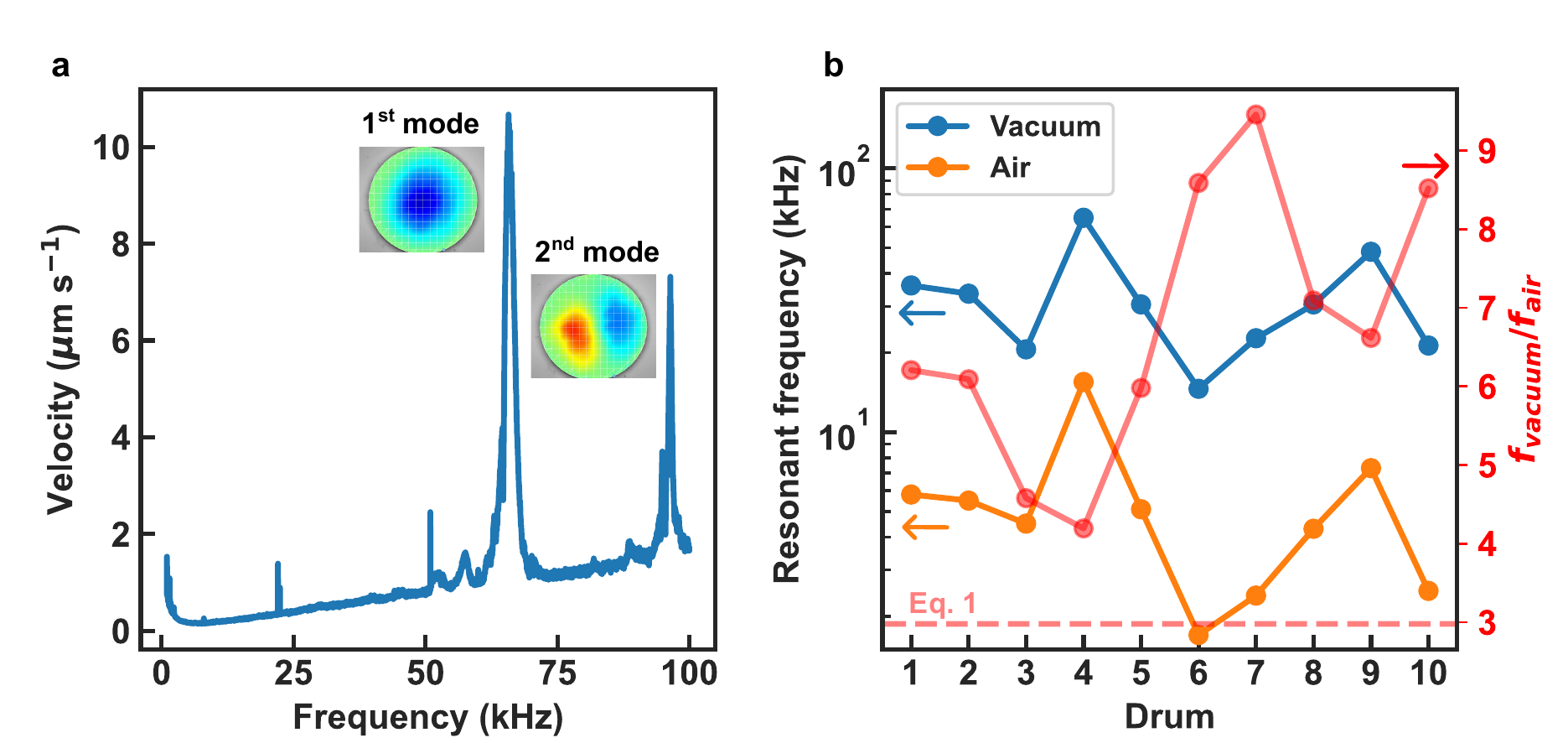}
    \caption{\textbf{Resonance frequency: vacuum vs air} (a) Mechanical spectra of a graphene membrane under piezoelectric actuation measured through a scanning laser vibrometer. Mode shapes corresponding to the first and second resonant peaks are also shown. (b) Comparison between resonance frequency measured in vacuum and in air (left y-axis) for different graphene drums. The ratio $f_\mathrm{vacuum}/f_\mathrm{air}$ along with the expected value from Eq. 1 are also shown (right y-axis). Added air mass causes a significant decrease of the mechanical resonance frequency which is not well reproduced by theoretical model.}
    \label{fig:mode}
\end{figure}

Apart from detecting the resonance frequency in air using acoustic actuation, we also investigated the resonant response in vacuum using a scanning laser vibrometer from Polytec. The sample was placed in a vacuum chamber ($10^{-3}$ mbar) on a piezo-shaker used to actuate the mechanical resonance. The mechanical response was detected over a grid of points distributed over the surface of the membrane such that the mode shape at resonance could be reproduced. Figure \ref{fig:mode}a shows a typical spectrum with the resulting shape of first and second resonant mode for a membrane with $d = 350\mu m$. In table S1, we show the resonance frequency $f_0$ and $Q$-factor resulting from fit of curves in Fig. 3a of the main text (atmospheric conditions) as well as from fit to data like in Fig. S4 (vacuum condition $10^{-3}$ mbar). The origin of the different values for atmospheric and vacuum condition is not understood yet.
\renewcommand{\thetable}{S1}
\begin{table}[H]
\centering
\begin{tabular}{ccccc}
Drum & $f_0$ (kHz) & $Q$ & $f_0$ (kHz) & $Q$\\\hline
1 & 1.5 & 1.1 & 14.4 & 161  \\
2 & 2.6 & 1.6 & 21.4 & 254 \\
3 & 4.6 & 1.5 & 22.7 & 268 \\
4 & 6.7 & 3.1 & 33.5 & 236 \\
\hline
\end{tabular}
\caption{\textbf{Resonance frequency and $Q$-factor of drums from Figure 3a.} Resonance frequency and $Q$-factor resulting from fit of curves in Fig. 3a of the main text (atmospheric conditions), where drums are numbered from the top curve to bottom curve in Fig. 3a. Resonance frequency and $Q$-factor of same drums from fit to data like in Fig. S4 (vacuum condition $10^{-3}$ mbar).}
\label{tableS1}
\end{table}
A comparison between resonance frequencies measured in vacuum and air is shown in Fig. \ref{fig:mode}b. The observed ratio between resonance frequency in vacuum and in air is $f_\text{vacuum}/f_\text{air} \sim 5-10$. The fundamental resonance frequency of a membrane can be expressed as $f_0 = \dfrac{2.405}{2\pi R}\sqrt{\dfrac{n_0}{m_\text{eff}}}$.
In atmospheric condition, the surrounding air is pushed along, adding mass to the resonator. In previous works \cite{al-mashaal_dynamic_2017, xu_realization_2021} the membrane's effective mass is taken as 
\begin{equation}
m_\text{eff} = \rho t\left(1 +\dfrac{2\rho_\text{air}R}{3\rho t}\right)\,,
\end{equation}
where $\rho$, $t$ and $R$ are the density, thickness and radius of the membrane and $\rho_\text{air}$ is the density of air. As shown in Fig. \ref{fig:mode}b, Eq. 1 predicts a ratio $f_\mathrm{vacuum}/f_\mathrm{air}$ of $\sim$ 3.5 whereas the experimental ratio is higher, between 4-10. However, Eq. 1 was developed for valveless micropumps in the plate limit \cite{pan_analysis_2003}, which can explain the disagreement between our measurement and Eq. 1. It might also be that the mass density of the graphene membrane $\rho t$ is  lower than that of crystalline graphene.  Further investigation is needed to understand better this effect on thin membranes dominated by in-plane pretension.

\section{S5. Wrinkles}
\begin{figure}[H]
    \centering
    \includegraphics[width = \linewidth]{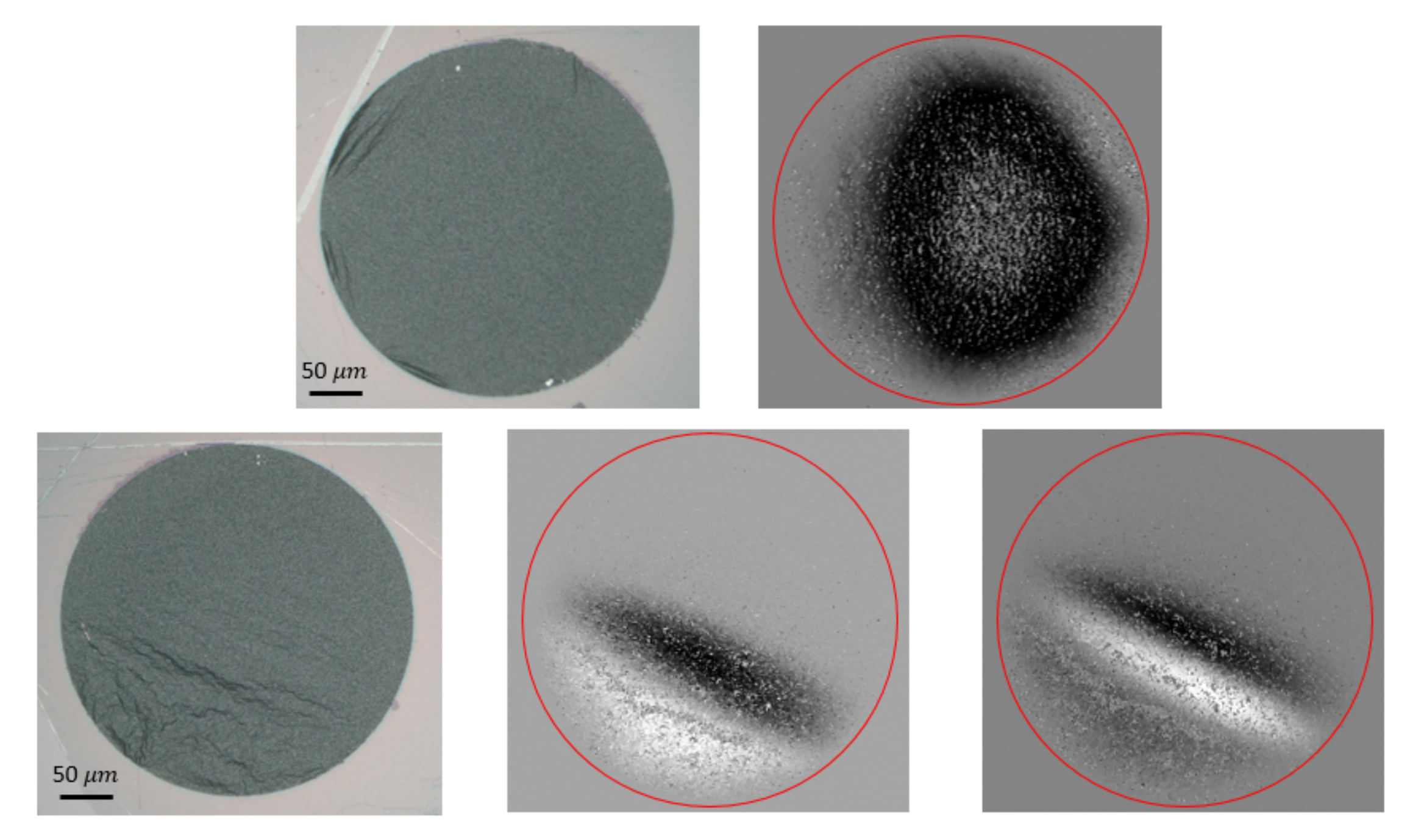}
    \caption{\textbf{Wrinkles and drum deformation} Examples of different deformation shapes on wrinkled graphene membranes measured with a stroboscopic technique in a DHM.}
    \label{fig:wrink}
\end{figure}

The sensitivity variation from sample to sample observed in Fig. 3b and S3 in the main manuscript was linked to the change in the membrane's pretension during the fabrication process. Moreover, wrinkles can form during the transfer process, introducing additional non-uniformities in the strain over the surface of the membrane which can vary between different samples. This phenomenon can lead to deviations from the ideal deformation shape of a membrane as shown in Fig. \ref{fig:mode}a and thus different mechanical response from the membranes. In Fig \ref{fig:wrink}, we show an example of two drums with different mode shape shapes arising from wrinkle-induced strain. The mode shapes were acquired with a digital holographic microscope (DHM) from Lynceetec paired with a laser pulsed stroboscopic unit. The measurements were carried out in vacuum conditions ($\sim 10^{-3}$ mbar) and the membranes were driven into resonance via piezoelectric actuation. The periodic signal driving the piezo-shaker was controlled via the stroboscopic unit to synchronize excitation signal with the DHM laser pulse and camera shutter.

Videos of the recorded mode shapes of two membranes are also included as supplementary material to further illustrate possible wrinkle-induces deformation of the mode shapes.

\section{S6. SEM picture of graphene membranes}
\begin{figure}[H]
    \centering
    \includegraphics[width = 0.495\linewidth]{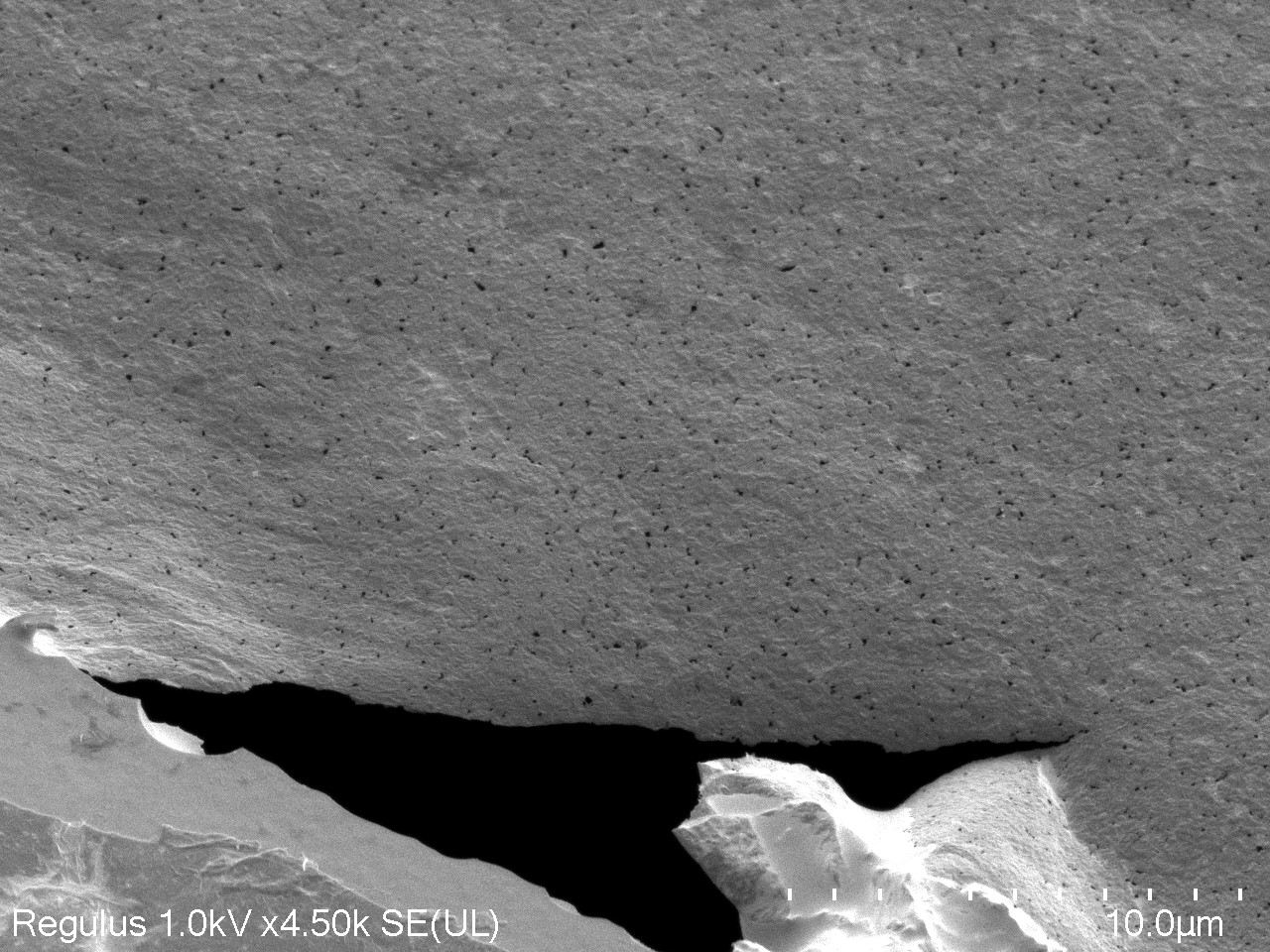}
    \includegraphics[width = 0.495\linewidth]{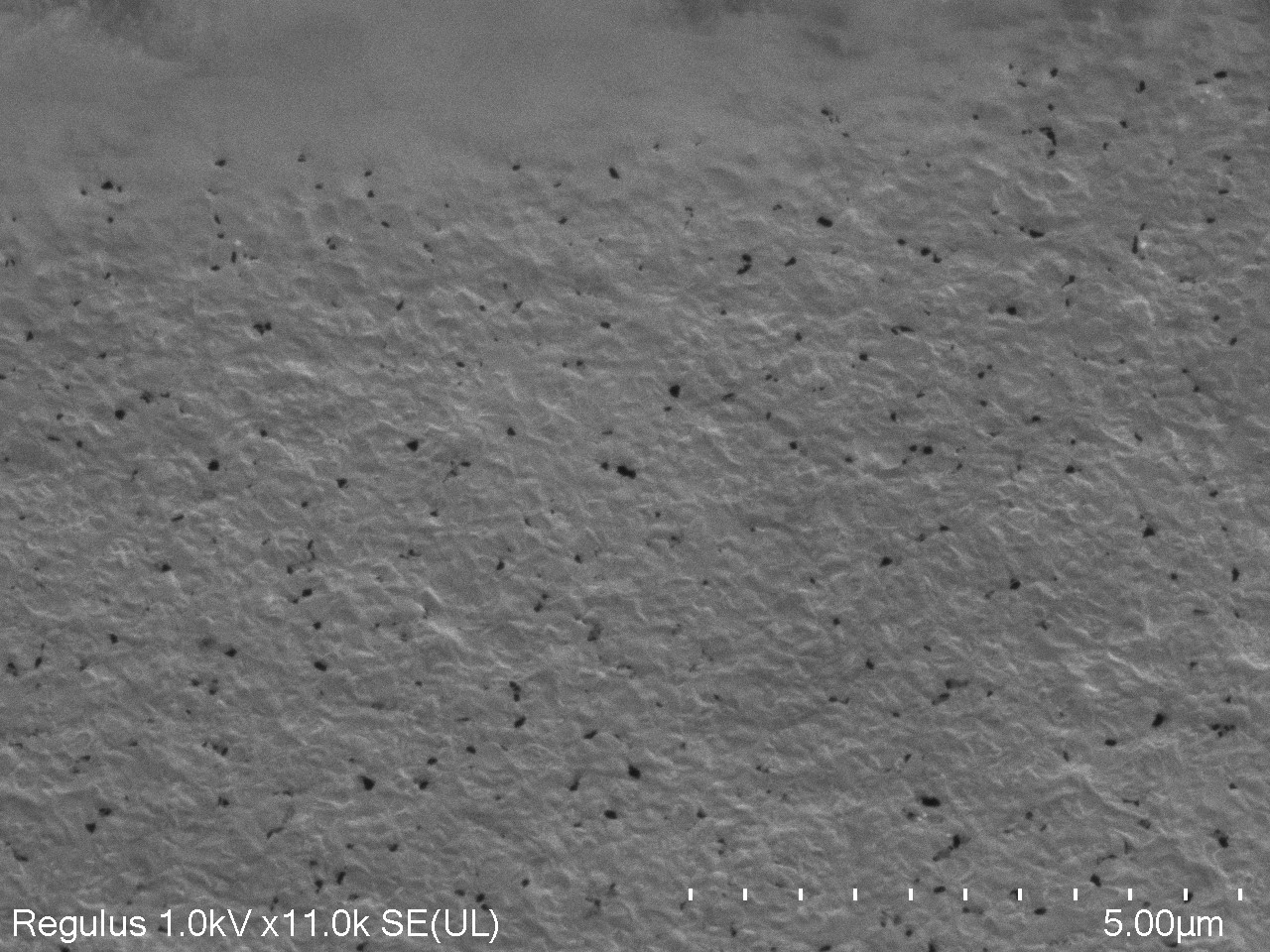}
    \includegraphics[width = 0.495\linewidth]{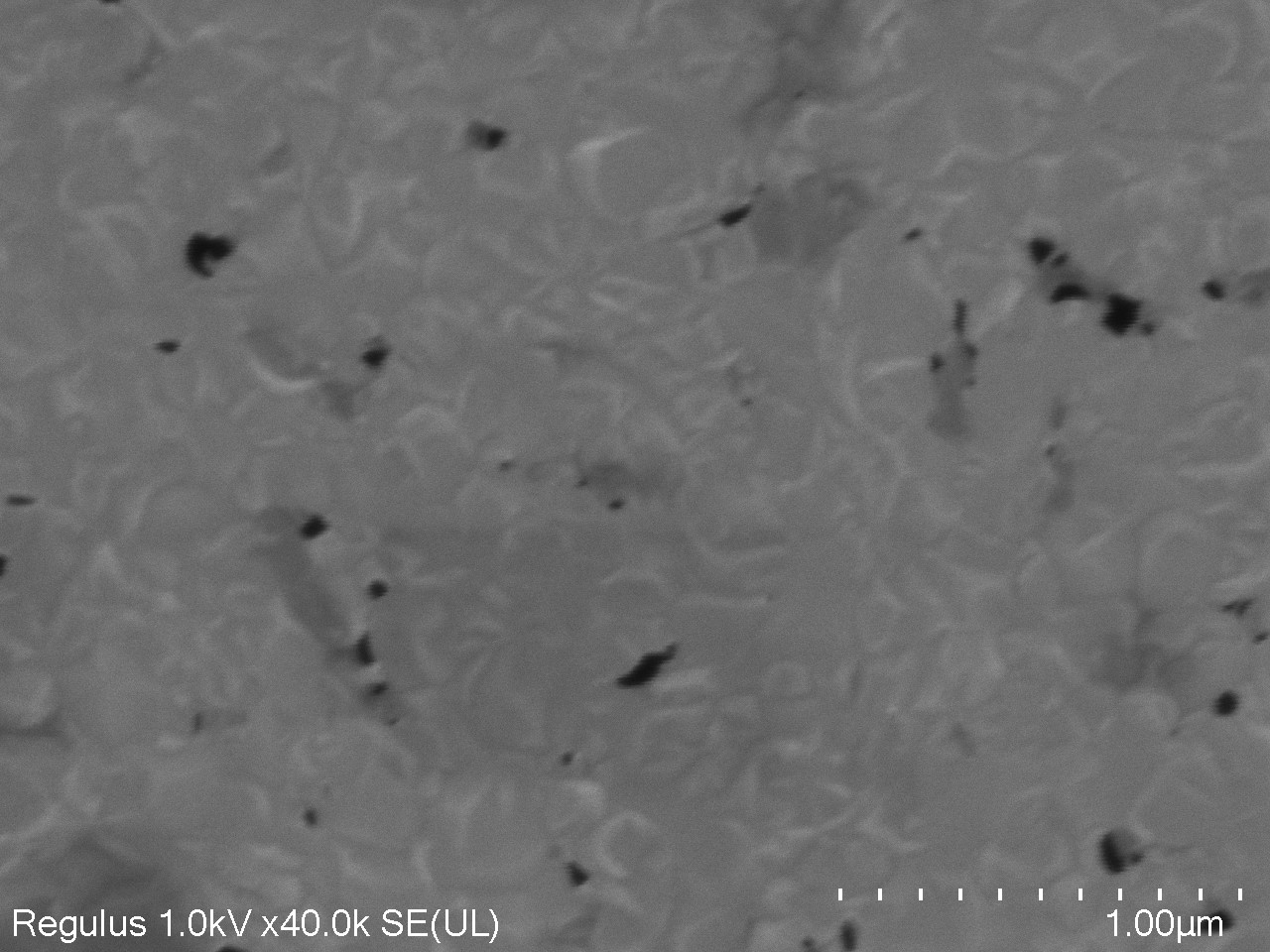}
    \caption{\textbf{SEM pictures of suspended membrane} SEM pictures of a broken suspended membrane showing small hole defects in the membrane.}
    \label{fig:sem}
\end{figure}

As mentioned in the main manuscript, small holes of $<$ 50 nm are left on the graphene during the CVD growth. The SEM pictures on Fig. \ref{fig:sem} show an example of these defects on a broken membrane.

\section{S7. Bandwidth correction to mechanical compliance}
As explained in the main text, we divided the mechanical compliances shown in Fig. 6 by the term (20 kHz$/f_0)^2$ to account for differences in the resonance frequency, $f_0$, between the different devices. We do this because if the tension $n_0$ of the membranes would be increased by a factor $(20 \text{kHz}/f_0)^2$  then their resonance frequency would increase by a factor $20 \text{kHz}/f_0$ (since $f_0 \propto \sqrt n_0$), but their compliance would drop by a factor $(20 \text{kHz}/f_0)^2$ (since $C_m \propto n_0$). In Fig. S7 we plot the corrected values for the membrane in this work (orange hexagons), graphene membranes in literature reporting the membrane's resonance frequency (purple hexagons), MEMS devices from \cite{zawawi_review_2020} (blue circles) and the commercial ST MEMS device (yellow star). The corrected values for the graphene membranes in this work are calculated from the data in Fig. S3. The datasheet for the ST MEMS \cite{STM} does not report explicitly the resonance frequency of the device but reports device performance only up to 10 kHz. We therefore assume that its resonance frequency is $\leq 20KHz$ and used a value of 20kHz for this calculation. Thus, even after taking the difference in resonance frequency into account, the graphene membranes in this work exhibit higher compliance than most devices reported in literature.

\begin{figure}[H]
    \centering
    \includegraphics[width = 0.6\linewidth]{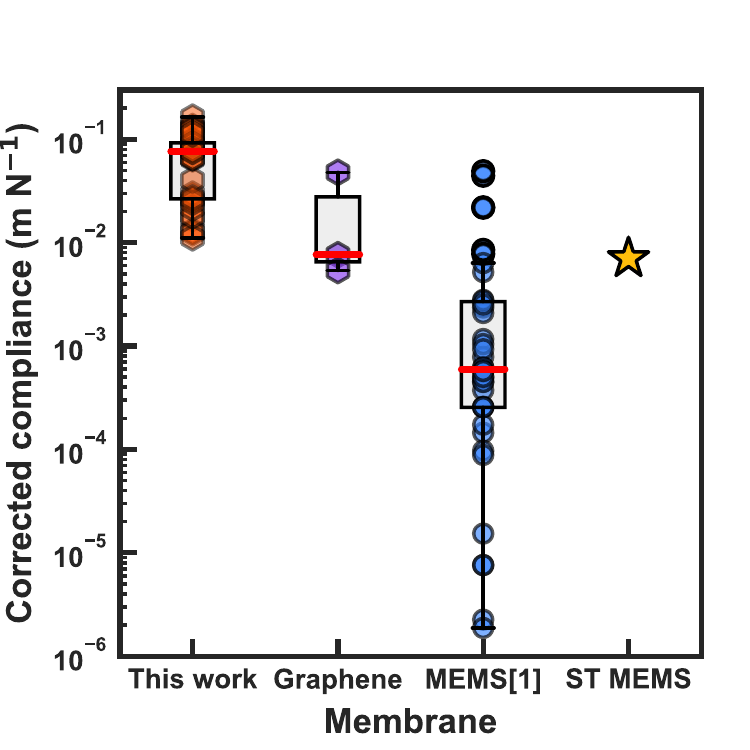}
    \caption{\textbf{Corrected compliance} Box plot of the mechanical compliance  of membranes studied in this work normalized by (20 kHz$/f_0)^2$.}
    \label{fig:compl}
\end{figure}

\section{S8. Sound recording using the graphene membrane}
In order to demonstrate the performance of the graphene membranes as microphones, we record a music soundtrack by optical readout of the graphene motion. The output signal from the vibrometer in response to sound was measured with a sampling frequency of 20 kHz. The recorded waveform was converted to back to an audio file which is included as supporting material.

\bibliography{main}

%\newpage
%\section*{For Table of Contents Only}
%\begin{figure}
%    \centering
%    \includegraphics{Figures/TOC.pdf}
%\end{figure}
\end{document}